\documentclass[conference]{IEEEtran}
\ifCLASSINFOpdf
  \usepackage[pdftex]{graphicx}
\else
  \usepackage[dvips]{graphicx}
\fi
\usepackage[cmex10]{amsmath}

\usepackage{stfloats}

\usepackage{bbm}

\newcommand{\norm}[1]{\left|\left|#1\right|\right|}
\newcommand{\mbf}[1]{\mathbf{#1}}

\begin{document}
%
\title{Finding role communities in directed networks using Role-Based
  Similarity, Markov Stability and the Relaxed Minimum Spanning Tree 
  }

\author{\IEEEauthorblockN{Mariano Beguerisse-D\'iaz}
\IEEEauthorblockA{Department of Mathematics\\
Imperial College London\\
London, SW7 2AZ\\
United Kingdom\\
m.beguerisse@imperial.ac.uk}
\and
\IEEEauthorblockN{Borislav Vangelov}
\IEEEauthorblockA{Department of Mathematics\\
Imperial College London\\
London, SW7 2AZ\\
United Kingdom\\
borislav.vangelov09@imperial.ac.uk}
\and
\IEEEauthorblockN{Mauricio Barahona}
\IEEEauthorblockA{Department of Mathematics\\
Imperial College London\\
London, SW7 2AZ\\
United Kingdom\\
m.barahona@imperial.ac.uk}}



\maketitle

\begin{abstract}
We present a framework to cluster nodes in directed networks according
to their roles by combining Role-Based Similarity (RBS) and Markov
Stability, two techniques based on flows. First we compute the RBS
matrix, which contains the pairwise similarities between nodes
according to the scaled number of in- and out-directed paths of
different lengths. The weighted RBS similarity matrix is then
transformed into an undirected similarity network using the Relaxed
Minimum-Spanning Tree (RMST) algorithm, which uses the geometric
structure of the RBS matrix to unblur the network, such that edges
between nodes with high, direct RBS are preserved. Finally, we
partition the RMST similarity network into role-communities of nodes
at all scales using Markov Stability to find a robust set of roles in
the network. We showcase our framework through a biological and a
man-made network.
\end{abstract}


%
\IEEEpeerreviewmaketitle

\section{Introduction}

Among the many systems that can be formalised as networks, there are
important examples where the directionality of the network is crucial,
e.g., the web, ecological systems, information and transport networks.
However, directionality brings in subtle mathematical complexities and
is often neglected in many approaches for network analysis. Such
directed networks naturally lend themselves to be analysed from the
perspective of \textit{flows}.  An important aspect of directed
networks is the notion of {\it roles}, e.g., leader vs follower or hub
vs authority. Here we show that a nuanced classification of nodes in
terms of their role in the network may be obtained from the analysis
of directed flows.  In other words, we seek to find nodes that are
similarly positioned in the network--with respect to flows--and obtain
broad categories into which they can be classified. In this paper, we
present a method to find role clusters in directed networks based on
the analysis of flow patterns, and show examples of its application to
a selection of networks.
Section~\ref{sec:theory} contains a brief introduction and references
to the specific techniques we use and an overview of the method.
Section~\ref{sec:examples} provides examples of our method.

\section{Methodology}
\label{sec:theory}

Let $\mathcal{G}= \{\mathcal{N}, \mathcal{E}\}$ be an unweighted and
directed network with node set $\mathcal{N}$, $|\mathcal{N}|=N$, edge
set $\mathcal{E}$, and adjacency matrix $\mbf{A}$ where $a_{i,j}=1$
denotes the existence of a directed edge from node $i$ to $j$. Each
node has in-degree $k_{in}$ (the number of nodes that
link to it) and out-degree $k_{out}$ (the number of nodes to
which it links). The $N\times 1$ vectors of in- and out-degrees are
denoted by $\mbf{k}_{in}$ and $\mbf{k}_{out}$. 

We find the role-communities of $\mathcal{G}$ following three steps:
\begin{enumerate}
  \item From the adjacency matrix $A$, construct a $N \times N$ node
    similarity matrix based on the directed connectivity profile of
    the nodes, as given by RBS (Sec.~\ref{sec:RBS}).
  \item From this RBS matrix, obtain a (new) undirected similarity
    network using the RMST algorithm such that two nodes are connected
    if their connectivity profiles are highly similar
    (Sec.~\ref{sec:RMST}).
  \item Find robust partitions of this RMST similarity network into
    communities of nodes with the same roles at several levels of
    resolution using Markov Stability, a multiscale community
    detection method (Sec.~\ref{sec:Stability}).
\end{enumerate}

\subsection{Role-Based Similarity in directed networks}
\label{sec:RBS}

In a directed network the in- and out-connectivities of the nodes
contain information about the role of each node in the network. The
simplest categorisation of nodes into ``leaders'' and ``followers'',
according to the predominance of their in- or out-degree, is often
illustrative but limited as it neglects the full topology and
complexity of the network. Other methods that harness further
information from the network structure can be used to compute the
``status'' index~\cite{Katz1953}, PageRank~\cite{Page1999}, or the
``Hub''/``Authority'' score~\cite{Kleinberg1999}. Though powerful,
these methods are limited by the fact that they split the nodes into
at most two categories (or further categories according to a
one-dimensional classification).

To go beyond the `leader-follower' dichotomy, we employ Role-Based
Similarity (RBS)~\cite{Cooper2010, Cooper2010a}, a method that
calculates how similar nodes are to each other in terms of the scaled
number of adjacent directed paths of {\it all meaningful lengths}
(i.e., no longer than $N$). The idea is to create a $1\times 2K_{max}$
feature vector for each node, $\mbf{x}_i$, whose entries are the
weighted number of paths of lengths from 1 to $K_{max} < N$
originating and ending in node $i$.  All the feature vectors
$\mbf{x}_i$ are stored as the rows of the $N\times 2K_{max}$ matrix:
\begin{equation} 
  \mbf{X}= \left[ \ldots, \left. \left(\beta \mbf{A}^T
      \right)^k\mathbbm{1}, \dots \right|
     \ldots , \left(\beta \mbf{A}
      \right)^k\mathbbm{1},   \ldots  \right],
  \label{eq:X}
\end{equation}
where $k=1,\dots, K_{max}$. Note that the number of originating paths
of length $k$ from all nodes is given by
$\left(\beta\mbf{A}\right)^k\mathbbm{1}$, and the number of arriving
paths $\left(\beta\mbf{A}^T\right)^k\mathbbm{1}$, where $\mathbbm{1}$
is the $N\times 1$ vector of ones.  Here, we use
$\beta=\alpha/\lambda_1$, where $\lambda_1$ is the largest eigenvalue
of $\mbf{A}$ and $\alpha\in (0,1)$, which assures convergence of the
sequence $\beta^k\mbf{A}^k$ as $k\to \infty$.  Hence the columns of
$\mbf{X}$ contain the number of in- (or out-) paths of length $k$ for
each node weighted by $\beta^k$.

In addition to guaranteeing convergence, the parameter $\alpha$ also
determines the weight given to each path length: smaller values of
$\alpha$ give more weight to shorter paths than to longer ones. If
$\alpha \ll 1$, the columns of $\mbf{X}$ converge rapidly (because
$\lim_{k\to\infty} \left(\beta\mbf{A}\right)^k\mathbbm{1}=\mbf{0}$),
which results in feature vectors based only on local properties based
on short paths (i.e., in the limit $\alpha \to 0$, the feature vector
only contains $k_{in}$ and $k_{out}$). As $\alpha$ is increased, we
incorporate more global features of the network in our analysis.
Results in~\cite{Cooper2010} indicate that $\alpha = 0.95$ provides a
good balance between the information gathered from the local and
global flow structure in the network. However, the systematic
determination of $\alpha$ for each network is currently the focus of
further investigation.

From $\mbf{X}$ we then compute the RBS matrix $\mbf{Y}$, whose entries
contain the cosine-similarity between all rows of $\mbf{X}$:
\begin{equation}
  y_{i,j} = \frac{\mbf{x}_i
    \mbf{x}_j^T}{\norm{\mbf{x}_i}_2\norm{\mbf{x}_j}_2}.
 \label{eq:RBS_sim}   
\end{equation}
When nodes $i$ and $j$ have an {\it identical} pattern of path flows
\emph{at all lengths} in $\mathcal{G}$, then $\mbf{x}_i$ and
$\mbf{x}_j$ are collinear and $y_{i,j} \simeq 1$. On the contrary,
when nodes $i$ and $j$ do not have any number of paths in common at
any length (e.g., when $i$ is a source and $j$ a sink node) then
$y_{i,j}=0$. The RBS matrix $\mbf{Y}$ is symmetric, usually full, and
could be used to find groups of nodes with similar
connectivity. However, as is usually the case with correlation or
distance matrices, clustering $\mbf{Y}$ directly is problematic
because of its lack of sparsity and the unstructured nature of
geometric distances in high-dimensional spaces.  To unblur the
structure of $\mbf{Y}$, we extract a similarity network that select
links between nodes with strong similarity while discarding weak
similarities that can be explained in terms of other relationships in
the network, as we explain now.

\subsection{Obtaining the similarity network from the RBS matrix}
\label{sec:RMST}

The $N$ feature vectors containing the flow profiles of the nodes are
defined in a high-dimensional space of $2K_{max}$ dimensions. However,
because the coordinates of the vectors are smoothly related to each
other, we expect that the vectors of all nodes will lie in a lower
dimensional manifold whose structrure can be well captured by a graph
with a geometric structure.
Here we use the Relaxed Minimum-Spanning Tree (RMST) algorithm, a method that
incorporates local and global features of the data to recover such a
network from the RBS matrix. 

First, we define the `dissimilarity' (or `distance') matrix $\mbf{Z}$,
with $z_{i,j} = 1-y_{i,j}$, i.e., the more similar $i$ and $j$ are to each other, 
the smaller the value of $z_{i,j}$ and the closer $i$ and $j$ lie.  The RMST algorithm
constructs a network with adjacency matrix $\mbf{E}$ from $\mbf{Z}$ as
follows. First, consider $\mbf{Z}$ to be the adjacency matrix of a
weigthed graph and obtain a Minimum Spanning Tree (MST) in it, setting
$e_{i,j}=1$ if nodes $i$ and $j$ are neighbours in the tree. Each node
pair $(i,j)$ is connected by a path (or sequence of edges) $\{(i,k),
(k,h), \dots, (m,j) \}$ in the MST. We then find the maximal weight in
$\mbf{Z}$ along the MST path:
$$\textrm{mlink}_{ij}=\max \{z_{i,k}, z_{k,h}, \dots, z_{m,j}\}.$$ If
$\textrm{mlink}_{ij}$ is significantly smaller than $z_{i,j}$ then the
MST-path is considered to be a good model to explain the similarity
between nodes $i$ and $j$ and discard the direct link between them,
i.e., we leave $e_{i,j}=0$.  If $z_{i,j}$ is comparable to
$\mathrm{mlink}_{ij}$ then there is not sufficient evidence to believe
that the MST-path is a better model and we include the direct link
$e_{i,j}=1.$ More precisely, we set $e_{i,j} = 1$ when
 \begin{equation}
   \textrm{mlink}_{ij} + \gamma(d_{i} + d_{j}) > z_{i,j},
\end{equation}
where $d_i = \underset{k}{\min}\,{z_{i,k}}$ and $\gamma$ is a
parameter ($\gamma=0.5$ here). The term $\gamma d_i$ approximates the
local distribution of points (in $\mbf{Z}$) around $i$ and is
motivated by the Perturbed Minimum Spanning Tree
algorithm~\cite{Carreira2005}.

The RMST similarity network is an unweighted, undirected graph where
two nodes are connected only if their flow feature vectors are highly
similar, \emph{regardless of whether they are neighbours in the
  original graph $\mathcal{G}$ or not}.  We can also obtain a weighted
similarity graph by Hadamard-multiplying $\mbf{E}$ and $\mbf{Y}$. The
RMST network is sparse if the data in $\mbf{Y}$ results from a local
geometric structure (which the RMST tries to recover), and is more
amenable to analysis using network analysis techniques such as the
community detection method we describe below.

\subsection{Role-communities with Markov Stability}
\label{sec:Stability}

\begin{figure*}[!t]
\centering
\includegraphics[width=6.8in]{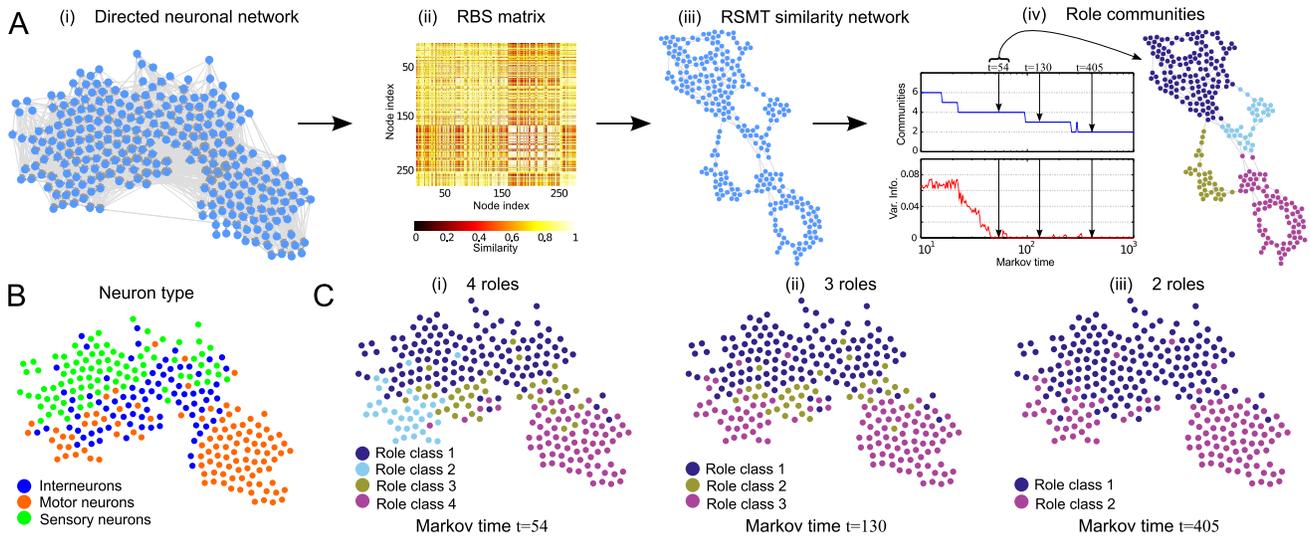}
\caption{Detecting role communities in the {\it
    C. elegans} neural network.  {\bf A}: {\it (i)}: Directed neural
  network~\cite{Varshney2011}. {\it (ii)}: Heatmap
  of the RBS matrix ($\mbf{Y}$, Sec.~\ref{sec:RBS}): the
  higher the similarity between two nodes, the lighter the cell. 
  {\it (iii)}: Similarity network obtained from the
  RBS matrix with the RMST algorithm (Sec.~\ref{sec:RMST}): only
  nodes with highly similar in- and out-flow patterns are connected. {\it (iv)}: Community detection of the RMST similarity network using Markov Stability (Sec.~\ref{sec:Stability}): robust partitions into 4, 3 and 2 role-communities are detected at different levels of resolution. 
  {\bf B}: Types of neurons in the {\it C. elegans} network, as given in Ref.~\cite{Varshney2011}. 
    {\bf C}: The role-classes obtained using 
    RBS+RSMT+Stability at different levels of resolution (Markov times).}
\label{fig:celegans}
\end{figure*}

Community detection in networks has been studied extensively and there
exist a wide variety of methods, each with their own
advantages~\cite{Fortunato2010}. Here, we use the method known as
Markov Stability~\cite{Delvenne2010, Delvenne2013} to detect
`role-communities' in the RMST similarity network.  There are a number
of advantages to using Markov Stability in this case, key among them
is the ability to detect communities in the network at {\it all
  scales} via a continous-time Markov process evolving in time.  Due
to this dynamic zooming, Markov Stability does not impose an \textit{a
  priori } number of roles (i.e., role-communities) in the network,
but rather detects the presence of robust partitions at all levels of
resolution. Hence we learn the number of roles by exploring the
network with a continuous-time diffusion process and finding robust,
optimised partitions across scales.

Consider $\mbf{E}$, the adjacency matrix of the undirected RMST
similarity network, which by construction is connected. Define
$\mbf{k}= \mbf{E}\mathbbm{1}$, the vector of degrees, and
$\mbf{D}=\mathrm{diag}(\mbf{k})$ so that $\mbf{D}^{-1}\mbf{E}$ is a
row-stochastic matrix. The normalised Laplacian of the system is then
$\mbf{L} = \mbf{I}_N - \mbf{D}^{-1}\mbf{E}$, and the transition matrix
of the continuous-time Markov process of duration $t>0$ (the Markov
time) is $P(t) = e^{-t\mbf{L}}$ (giving the probability of
transitioning form node $i$ to $j$ in a process of duration
$t$)~\cite{Lambiotte2008}.  A hard partition of the network into $C$
groups of nodes can be encoded in a $N\times C$ matrix $\mbf{H}$
(i.e., $\mbf{H}_{i,c}=1$ is node $i$ belongs to community $c$, and
$\sum_c{\mbf{H}_{i,c}}=1\,\forall\, i$). The {\it Markov Stability} of
the partition at time $t$ is defined as the trace of the clustered
autocovariance of the diffusion process~\cite{Delvenne2010}:
\begin{equation}
  r(t, \mbf{H}) = \mathrm{trace}\left(\mbf{H}^T\left[
  \Pi P(t) - \pi\pi^T\right]\mbf{H}\right),
  \label{eq:stability}
\end{equation}
where $\pi$ is the steady-state distribution of the process, and
$\Pi=\mathrm{diag}({\pi})$. We find the communities in the similarity
network for a given $t$ by maximising $r(t, \mbf{H})$ over the space
of partitions; that is, we find the network partitions that maximise
the retention of flows over a timescale. 

Maximising equation~(\ref{eq:stability}) is an NP-hard problem, with
no guarantees of global optimality. The optimised partitions are found
using the Louvain algorithm~\cite{Blondel2008}, a greedy heuristic
that has been shown to give good results in practice. To find the
relevant partitions of the network at any time scale, we use a
robustness criterion based on the consistency of the optimisation
quantified through an information-theoretical measure.  At each Markov
time, we obtain optimised partitions of the network by running the
Louvain method 100 times, each time using a random initial guess. To
gauge the robustness of the set of optimised partitions, we calculate
the mean pairwise Variation of Information (VI) of the ensemble of
Louvain solutions~\cite{Delvenne2013}.  The VI between two partitions
$\mbf{H}_1$ and $\mbf{H}_2$ is~\cite{Meila2007}:
$$VI(\mbf{H}_1,\mbf{H}_2)=\frac{1}{\log{N}}\left(2H(\mbf{H}_1,\mbf{H}_2)-
H(\mbf{H}_1)-H(\mbf{H}_2)\right),$$ with
$H(\mbf{H})=-\sum_c{p(c)\log{p(c)}} \text{ and }
p(c)=\sum_i{\mbf{H}(i,c)}/N$.  When the optimisation algorithm finds
partitions of the network that are consistently similar (a hallmark of
robust community structure), the mean pairwise VI is low; when there
is no clear community structure the optimisation produces partitions
that are different to each other, resulting in a high mean
VI. Finally, to make sure we detect all the relevant role-communities,
we optimise equation~(\ref{eq:stability}) for all Markov times,
keeping track of the mean VI as a function of $t$, and detecting
communities that are also persistent across Markov times.  We now
provide examples of finding node roles in different networks using the
RBS/RMST/Markov Stability methodology explained in this section.

\section{Examples}
\label{sec:examples}

\subsection{C. elegans neural network}
\label{sec:celegans}

The directed neural network of {\it C. elegans} records the chemical
synapses and the junctions between 279 neurons~\cite{Varshney2011}.
Figure~\ref{fig:celegans}A shows the steps of the analysis: the
original {\it unweighted} directed neuronal network; computation of
the RBS matrix $\mbf{Y}$ with $\alpha=0.95$ and $K_{max}=116$;
generate the RMST similarity network; and find role-communities in it
using Markov Stability.  As shown in Fig.~\ref{fig:celegans}A-(iv),
our analysis finds meaningful partitions of the RMST network into up
to four role-classes.

{\it C. elegans} is known to have three types of neurons: sensory,
motor, and interneurons~\cite{Varshney2011}, shown in
Figure~\ref{fig:celegans}B with different colours. We display the
neural network on the plane as in Ref.~\cite{Varshney2011}: the
horizontal axis corresponds to the entries of the Fiedler vector
reflecting mostly body position, and the vertical axis corresponds to
processing depth with respect to information flow.
Fig.~\ref{fig:celegans}C shows that the partition into four, three and
two roles broadly reflects the biological groups.  Among the four
roles, we find one group (Role 1, dark
blue) which corresponds mostly to sensory neurons and some
interneurons. Role 3 is formed by a subset of the interneurons.
Interestingly, motor neurons are split in two classes (Roles 2 and 4)
clearly separated along the body of the worm (x-axis).  The motor
neurons in Roles 2 and 4 are merged into a common role when we cluster
the RMST network into 3 role-communities. Finally, the partition
into two roles separates the groups along the lines of sensory and motor
neurons---interneurons are split in both, with slightly more interneurons in the
sensory group.

\subsection{US airport network}
\label{sec:usairports}

\begin{figure}[!t]
  \centering
  \includegraphics[width=3.45in]{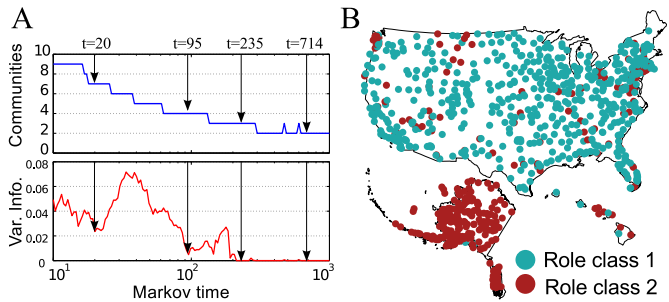}
  \caption{Roles in the US airport network. 
  {\bf A}: Number of role-communities found in the RMST similarity
    network at all Markov times (top) and the Variation of Information of
    the partitions (bottom).  {\bf B}: The two roles found in the US
    airport network at Markov time $t=714$.}
  \label{fig:usairports}
\end{figure}

We also investigated the roles in the unweighted network of $N=957$
airports in the United States ~\cite{Opsahl2011, Kunegis2012}.  The
analysis proceeds as before and we calculate the RBS matrix with
$\alpha=0.92$ and $K_{max}=78$.  Figure~\ref{fig:usairports}A shows
the number of roles found at different levels of resolution.  We find
partitions into seven or fewer role classes---the VI has pronounced
dips at $t=20$, $t=95$, $t=235$, and $t=714$, corresponding to 7, 4, 3
and 2 role-communities. Interestingly, there is always a distinctive
role for a large group of Alaskan airports across all levels of
resolution which persists separately up to the highest level of resolution. As shown in
Fig.~\ref{fig:usairports}B, where we show the US map with nodes
coloured according to the two role classes at $t=714$, the most
striking attribute is that practically all airports
in Alaska (including the two largest, Anchorage and Fairbanks) belong
to role class 2.  Transport in Alaska, a large and sparsely populated
region with many remote settlements scantily connected by roads,
relies on local airports and airstrips. These ingredients
contribute to create a distinct (and less reciprocal) air-transportation
connectivity, which sets Alaska apart from most of the rest of the US.
The few nodes in the mainland and Hawaii that belong to role class 2
are mostly small airfields and industrial airports, which are embedded in local
air route patterns.

\section{Conclusion}

We show how directed flow patterns at all scales in directed networks
can be harnessed using a combination of flow-based and structural
approaches to uncover the different types of nodes that exist in a
directed network. Both RBS and Markov Stability at their core rely on
flows but each from a different stance: the former compares how the
similarly nodes are positioned with respect to incoming and outgoing
flows, while the latter establishes where flows tend to be trapped on
a given timescale. The RSMT algorithm allows us to project complex
datasets with local structure as true networks, facilitating its
analysis with graph theoretical tools.  Together, these techniques
form a powerful framework for the analysis of directed networks which,
as the examples here show, is applicable to networks originating from
different disciplines.


\section*{Acknowledgments}
MBD and MB acknowledge support from the UK EPSRC grant
EP/I017267/1 under the Mathematics Underpinning the Digital Economy
program. MBD acknowledges support from the James S. McDonnell
Foundation Postdoctoral Program in Complexity Science/Complex
Systems-Fellowship Award (\#220020349-CS/PD Fellow). BV is supported
through a PhD Award from the British Heart Foundation
Centre of Research Excellence at Imperial College London (RE/08/002).




\bibliographystyle{IEEEtran}

%

\end{document}